\documentclass[12pt]{iopart}
\usepackage{graphicx}
 \begin{document}
\title{Uncertainty study of nuclear model parameters for the n+$^{56}$Fe
reactions in the fast neutron region below 20 MeV}

\author{Junfeng Duan$^{1, \dag}$, Stephan Pomp$^{1, \dag}$, Henrik Sj{\"o}strand$^{1}$, Erwin Alhassan$^{1}$, Cecilia Gustavsson$^{1}$, Michael {\"O}sterlund$^{1}$, Arjan Koning$^{2}$, Dimitri Rochman$^{2}$}
 \address{$^{1}$Division of applied nuclear physics, Department of physics and astronomy, Uppsala University, Box 516, 751 20, Uppsala, Sweden\\ $^{2}$Nuclear Research and consultancy Group(NRG),P.O. Box 25, 3 Westerduinweg, 1755 ZG Petten, The Netherlands}
\ead{$^\dag$junfeng.duan@physics.uu.se; stephan.pomp@physics.uu.se}

\begin{abstract}  In this work, we study the uncertainty of nuclear model parameters for neutron induced $^{56}$Fe reactions in fast neutron region by using the Total Monte Carlo method. We perform a large number of TALYS runs and compare the calculated results with the experimental data of the cross sections to obtain the uncertainties of the model parameters. Based on the derived uncertainties another 1000 TALYS runs have been performed to create random cross section files. For comparison with the experimental data we calculate a weighted $\chi^2$ value for each random file as well as the ENDF/B-VII.1, JEFF3.1, JENDL4.0 and CENDL3.1 data libraries. Furthermore, we investigate the optical model parameters correlation obtained by way of this procedure.
\end{abstract}
\maketitle

\section{Introduction} Since the beginning of the century, the nuclear science community is putting more and more attention to the assessment of uncertainties which is important for both basic physics and technological applications. More recently, a new method has been developed and applied based on Monte Carlo calculations and called ``Total Monte Carlo", or TMC\cite{08koning}. This method relies on a large number of nuclear model calculations, with different random nuclear data in each of them. The essential idea of this method is to assume that each nuclear model parameter has its own uncertainty. Model calculation will be performed many times, whereby each time all elements of the input parameter vector are randomly sampled from a normal or uniform distribution,centered around a initial value, with a specific width for each parameter. At NRG, this method has been implemented in a code called TASMAN\cite{12koning}. In TASMAN, a binary reject/accept method is used to introduce correlations between the nuclear model parameters: calculations are judged by their agreement with experimental data(see sect. II A) and a TALYS\cite{07koning} run with a certain parameter set might be either accepted or rejected. In this way only certain model parameter combinations survive,and the parameter distribution is automatically determined numerically by experimental data, without having to resort to a priori distribution.

By using this method we investigate the uncertainty of model parameter and parameter correlations for n+$^{56}$Fe reactions. Based on the derived parameter uncertainties another a set of 1000 random cross section files has been created with TALYS. For comparison with the experimental data we calculate a weighted $\chi^2$ value for each random file as well as the existing evaluated data libraries, such as ENDF/B-VII.1, JEFF3.1, JENDL4.0 and CENDL3.1. Section II describes the method in more detail and gives the main results. A briefly summary is given in last section.

\section{Methodology and results}
\subsection{Methodology}
We use the following scheme in our work.
\begin{itemize}
\item Using the TALYS default parameters as starting point a manual search for a good set of model parameters that well describe experimental data in the considered energy region is performed. Note that resonances are not included in this work. All experimental data used in this work are taken directly from the EXFOR database.
\item For the next step we make a rough assumption on parameter uncertainties. These are used to sample parameter space in a number of TALYS runs. Given typical experimental uncertainties a TALYS run is accepted if it does not differ from the average result by more than the experimental uncertainty, and it is rejected otherwise. In this way we find both improved values for parameter uncertainties (given in Table \ref{Table1}) and obtain correlations between model parameters (see Sect. II.C).
\item Finally, using the derived parameter uncertainties, a set of 1000 randomized cross section files is obtained from TALYS
    and this set is directly compared to experimental data by means of a $\chi^2$ test as described below.
\end{itemize}

The run with the lowest $\chi^2$ value together with the parameter uncertainties derived in the fashion described above are a good starting point for use in TMC.
\begin{table}
\centering
\caption{Uncertainty of some nuclear model parameters for $^{56}Fe$, given by fraction(\%) of the absolute value.}
\label{Table1}
\begin{tabular}{cc|cc}  

\hline\hline
Parameter  &  Uncertainty(\%)  &Parameter  &  Uncertainty(\%)    \\
\hline
$r^n_V$       &  2 &   $a^n_V$ &  2.5   \\
 $v^n_1$ & 2& $v^n_2$     &  5\\
  $v^n_3$ &  9     & $w^n_1$ & 5\\
  $w^n_2$     &  7 & $d^n_1$ &  9.4    \\
 $d^n_2$ & 10 &$d^n_3$     &9.7 \\
    $r^n_D$ &  2.5   & $a^n_D$   & 3 \\
$r^n_{SO}$    &9.7 &$a^n_{SO}$ &  10    \\
 $v^n_{so1}$ & 15&$v^n_{so2}$ & 20 \\
 $w^n_{so1}$&  30  &$w^n_{so2}$& 35\\
 $r^p_V$ &  2   & $a^p_V$& 2 \\
$r^{\alpha}_V$ &  2 &$a^{\alpha}_V$ &  2   \\
$M_2$  &15 &$\Gamma_\gamma$&  8 \\
$a(^{57}Fe)$ &  4.5   &$a(^{56}Fe)$& 6.5\\
$a(^{55}Fe)$&  5 &$a(^{56}Mn)$&  5   \\
 $a(^{53}Cr)$ & 5 &$\sigma^2$  &  15\\
 $g_\pi$($^{57}Fe$) &  6.5   & $g_\nu$($^{57}Fe$)&8\\
$C_{knock}$&4 &$C_{strip}$&4\\
\hline\hline
\end{tabular}
\end{table}

\subsection{${\chi}^2$ calculation}
we calculate $\chi^2$ values for comparison with experimental data. First, the average $\chi^2_c$ for each reaction channel is calculated as follows
\begin{equation}
\chi^2_c={\frac{1}{N}}\sum_i^N{(\frac{\sigma_T^i-\sigma_E^i}{\Delta\sigma_E^i})}^2.
\label{chic}
\end{equation}
 Where $N$ is the number of experimental data of each reaction channel, $\sigma_T$ the calculated result, $\sigma_E$ the experimental data, $\Delta\sigma_E$ the uncertainty of the experimental data. The total average $\chi^2$ for each random file is defined as a sum over all reaction channels $N_c$
\begin{equation}
\chi^2=\frac{\sum_i^{N_c}w_i\chi^2_{ci}}{\sum_i^{N_c}w_i}.
\label{chi}
\end{equation}
We weight the $\chi^2$ with
\begin{equation}
w_i=\sqrt{\frac{1}{\Delta\overline{\sigma_E}}N\sigma_{max}}.
\label{w}
\end{equation}
 Here $\Delta\overline{\sigma_E}$  is the average uncertainty of each reaction channel, $\sigma_{max}$ the maximum value of the cross section. A large weight has been given for the reaction channel with a large number of experimental data, more precise data and large cross section. By this way, the total, elastic, nonelastic, inelastic reaction with large number data and large cross section, play the most important roles in $\chi^2$ calculation(see Table \ref{Table2}).

\begin{table}
\centering
\caption{List of $^{56}Fe$ experimental data selected for $\chi^2$ calculation with their weight.}
\label{Table2}
\begin{tabular}{cc|cc}  

\hline\hline
Reaction channel  &  Energy points  & Energy region(MeV)  &  weight     \\
\hline
(n,tot)&34&8-20&49\\
(n,el)&18&2-14&19\\
(n,non)&29&2-20&21\\
(n,$\gamma$)&1&14.2&0.04\\
(n,inl)&11&2-14.6&10\\
(n,$n^{\prime}_1$)&62&2-15.2&17.6\\
(n,$n^{\prime}_2$)&44&2.1-7.55&6\\
(n,$n^{\prime}_3$)&3&2.96-3.75&1.5\\
(n,$n^{\prime}_4$)&35&3.36-14.6&4.6\\
(n,$n^{\prime}_6$)&4&3.68-5.59&1.4\\
(n,2n)&27&11.88-19.3&11.6\\
(n,p)&300&5.46-20&4.5\\
(n,$\alpha$)&10&5.93-14.1&0.95\\
\hline\hline
\end{tabular}
\end{table}

\begin{figure}[t]
\includegraphics[width=0.95\columnwidth]{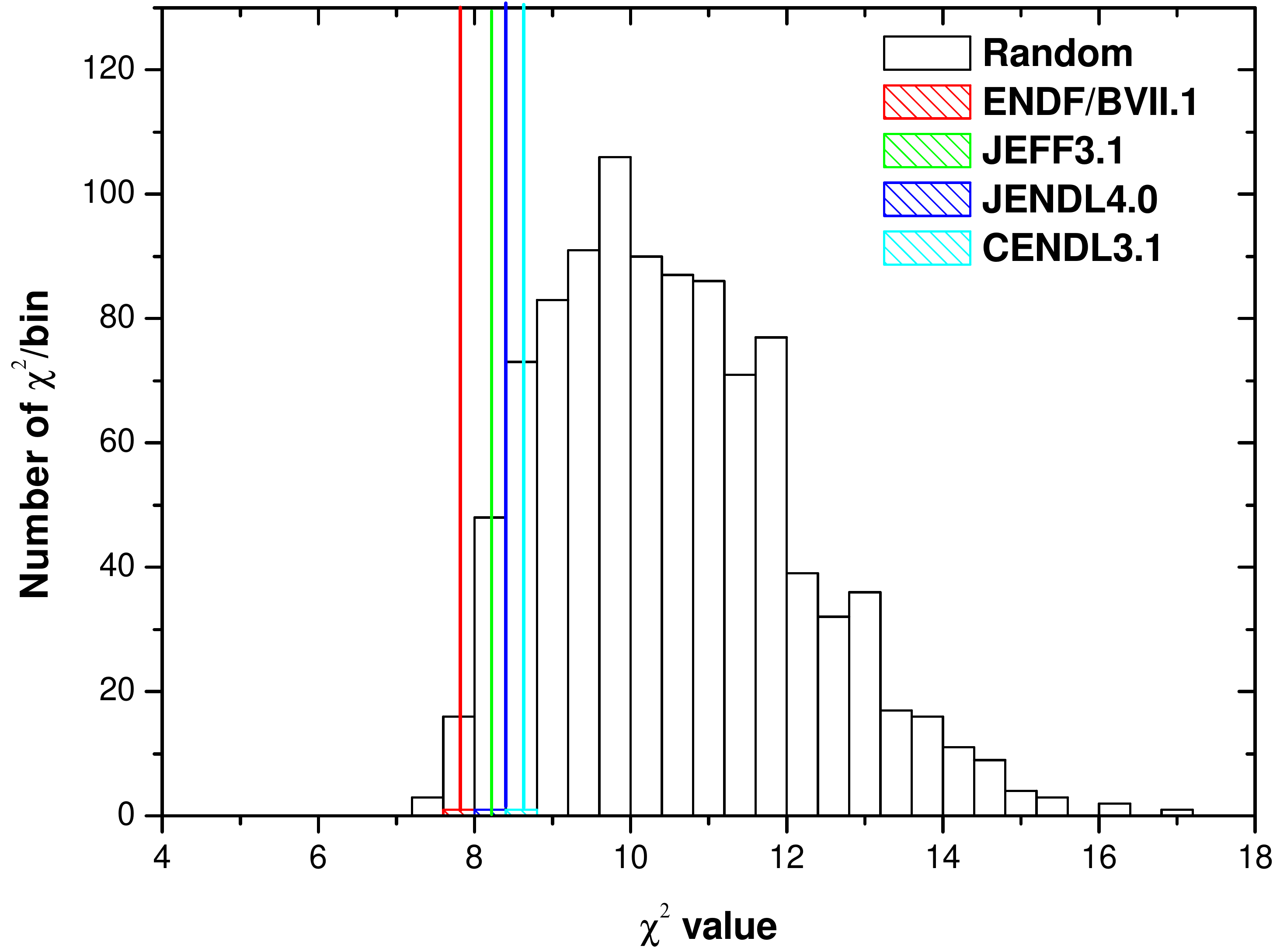}
\caption{$\chi^2$ values distribution for the random $^{56}Fe$ files per bin, compared to the $\chi^2$ values for the ENDF/B-VII.1, JEFF3.1, JENDL4.0 and CENDL3.1 libraries marked by the vertical lines with the corresponding color showed in figure.}
\label{chi2his}
\end{figure}
 Fig \ref{chi2his} presents the results of 1000 random $^{56}Fe$ files in terms of $\chi^2$ as defined in Eq.\ref{chi}, where each random $\chi^2$ is represented by an entry into the histograms. We also compare to the existing evaluations including ENDF/B-VII.1, JEFF3.1, JENDL4.0 and CENDL3.1. It can be seen from the figure that a number of our random cross section files perform as well as and in some cases even better than the evaluated libraries. As can be seen, the distribution is not symmetric, and has a tail towards high $\chi^2$ values\cite{12rochman}.

\subsection{Parameter correlations}
\begin{figure}[h]
\includegraphics[width=0.95\columnwidth]{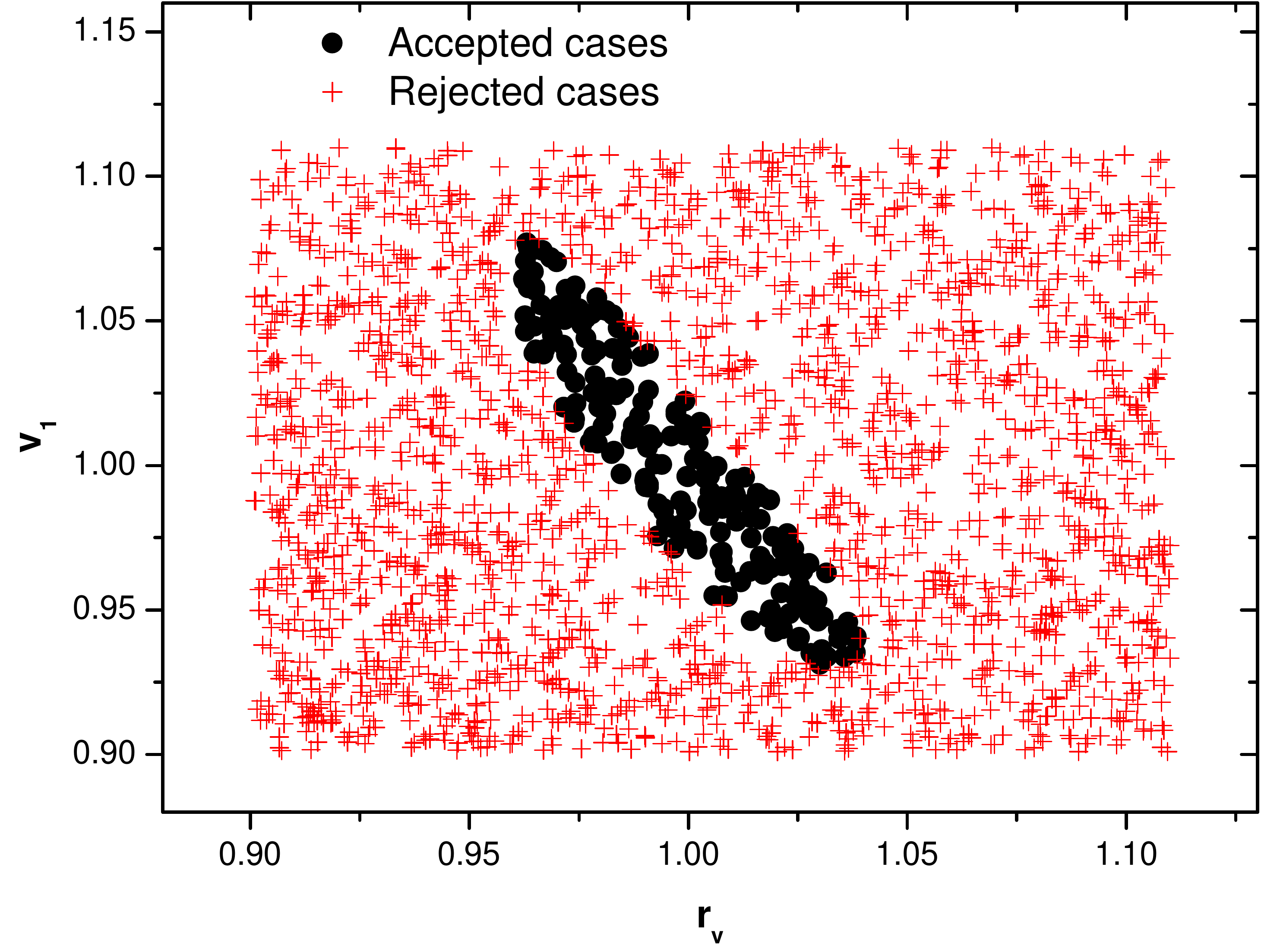}
\caption{Accepted and rejected cases decided by $r_v$ against $v_1$}
\label{rv_v1}
\end{figure}

\begin{figure}[h]
\includegraphics[width=0.95\columnwidth]{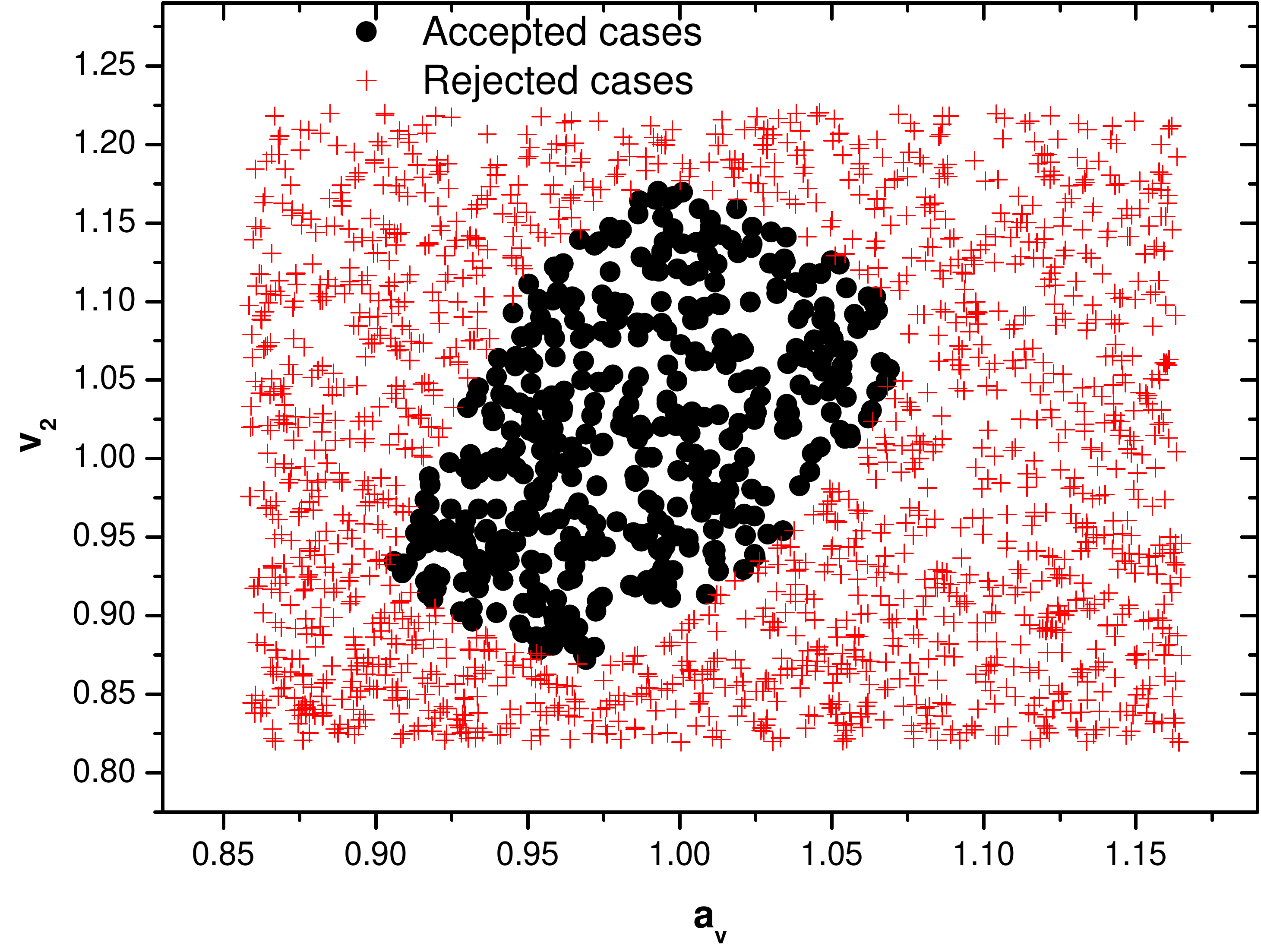}
\caption{Accepted and rejected cases decided by $a_v$ against $v_2$.}
\label{av_v2}
\end{figure}
The parameter correlations have been investigated by using the TASMAN code. In this case, we select any two of optical model parameters to sample with large uncertainty ranges while keeping the other parameters constant. Fig \ref{rv_v1} shows the accepted and rejected cases decided by $r_v$ against $v_1$. We use the larger uncertainties in this study in order to show the boundary of the accepted region. Considering the relation between $r_v$ and $v_1$, which is $v_1*r^2_v=constant$, $r_v$ and $v_1$ show the expected negative correlation. Actually, most optical model parameters show similar behaviour for the cross sections such as $r_v$ against $a_v$ and $d_1$ against $a_d$ etc, which have negative correlations. Fig \ref{av_v2} shows corresponding results for $a_v$ against $v_2$, and indicates that $a_v$ and $v_2$ have a positive correlation, i.e. $a_v$ and $v_2$ have opposite effect on the cross section. It can be seen from the figures that the slope of the accepted area is a measure for the parameter correlations.

\section{Summary}
In this work we obtain a set of uncertainties for the most important model parameters for n+$^{56}Fe$ reactions. 1000 TALYS runs have been performed to create random cross section files based on the derived uncertainties. For comparison with the experimental data, we calculate $\chi^2$ values with different weights for different reaction channels. Larger weights have been given for the reactions with a large number of experimental data, more precise data and large cross section. Judging from the obtained $\chi^2$, a number of random files perform as well as the evaluated libraries. Parameter correlation have been investigated by using TASMAN code. The results show correlations between optical model parameters. Meanwhile, this study further verified the reliability of the TASMAN code and the described reject/accept method. Actually, the whole TMC method include reactor code simulations, i.e. bechmark testing process. We also use TMC method for reactor code simulations on burn-up\cite{13henrik} and safety parameters\cite{13erwin}. Next step we will produce the ENDF random files including resonance part by using the whole TASMAN package for reactor code simulation.
\\

\end{document}